**Hole-doping driven antiparallel magnetic order underlying the superconducting and pseudogap states in high-temperature cuprate superconductor**

Takashi Uchino

Department of Chemistry, Graduate School of Science, Kobe University, Nada, Kobe 657-8501, Japan

Unveiling the nature of the pseudogap and its relation to both superconductivity and antiferromagnetic Mott insulators, the pairing mechanism, and a non-Fermi liquid phase is a key issue for understanding high temperature superconductivity in cuprates. A number of experimental findings gathered especially in recently years have revealed the coexistence of different types of electronic order, including charge- and spin-density-wave orders and nematic charge order. We here show that antiparallel magnetic order can be reasonably and naturally predicted in hole-doped $CuO_2$ planes by starting from the ground state of a weakly doped antiferromagnetic insulator, where a Skyrmion-type three-dimensional spin texture is created around the doped hole. The superconducting transition temperature $T_c$ can be understood in terms of the temperature at which long-range antiparallel magnetic ordering is established, resulting in the magnetically mediated superconducting state with phase-coherent Cooper pairs. Upon heating above $T_c$, long-range phase coherence in the pair state is lost, but the pair condensate still survives on the medium-range length scale, transforming to the pseudogap state with charge and magnetic orders. The present model further predicts that the CuO$_2$ plane with the hole concentration per Cu atom of $p$=1/6 (1/8) is at best (at worst) compatible with the hole-driven antiparallel magnetic order, rationalizing the existence of the optimally doped region and the so-called 1/8 anomaly. We believe the present model provides a key initial point to unravel a wide variety of the apparently complex phenomena related to high temperature cuprate superconductors.

Recent experiments in cuprate superconductors have revealed persistent evidence of charge- and spin-density-wave orders[1,2]. Thus, a complete theory of high-temperature superconductivity requires a full understanding of the temperature-dependent evolution of the spin-, charge, and orbital correlations. It is most likely that the copper-oxygen layer in doped cuprates plays a central role in exhibiting superconductivity and its relevant ordering states. However, a full understanding of the physics of a doped CuO$_2$ layer has not been achieved yet because of the presence of the strong electron-electron correlations in the layer[2-6]. This is the main reason why we still do not have a complete theoretical framework to understand the nature of high temperature superconductors. On the other hand, in undoped or slightly doped antiferromagnets, where an effective field theory for magnons and holes[7] is applicable, rather precise numerical simulations have been performed, as will be shown below. Thus, in this work, we first survey the present knowledge about a single hole in a CuO$_2$ plane. This is a crucial step to understand the change in the electronic structure of the $CuO_2$ plane with hole concentration.

**A single hole state in a $CuO_2$ plane**

In the undoped state, a $CuO_4$ plaquette in a copper-oxygen plane has a single spin residing mainly in the $3d_{x^2-y^2}$ orbital of the central copper site ($3d^9$). On doping, as has been demonstrated by Zhan and Rice[8] using a single-band effective Hamiltonian formalism, the additional hole is primarily located on the oxygen sites (Cu $3d^9$ O $2p^5$) and much less on the copper site (Cu $3d^8$) because of the strong Coulomb repulsion within the Cu $3d$ shell. Thus, the empty states near the Fermi energy tend to have O $2p$ character with increasing dopant concentration[9-11]. It is generally believed that the resulting $O^-$ hole locally forms a singlet state with the Cu spins and hence that the total spin becomes zero[8] (see Fig. 1a). This state, often called the Zhan-Rice singlet state, may be correct in the zeroth-order treatment indeed; however, it has now been recognized that the above simple singlet model should be modified at least by adding a hopping term to give a more realistic picture[12-20]. The addition of the hopping term yields a nonzero total spin[14,15], resulting in a Skyrmion-type topological spin texture with a core size of a few lattice spacing $a_0$[12,16-20], where $a_0$ is the distance of the Cu-O-Cu bond (see Fig. 1b). This indicates that the single-hole doping reduces the spin stiffness of the $CuO_2$ plane around the hole-doped region. More specifically, antiferromagnetic order in the $CuO_2$ plane is locally destroyed and is converted into a Skyrmion-type ferromagnetic order[19]. It has also been demonstrated that the doped hole localized around the central Cu site induces polaronic distortion[14,21,22]. Thus, a single hole doped in the $CuO_2$ plane can be best viewed as a small polaron with staggered magnetization order characteristic of a Skyrmion. In harmony with the scheme, it was recently inferred from low-temperature magnetic and transport measurements that Skyrmion-like ferromagnetic orders exist in a doped antiferromagnet ($La_2Cu_{0.97}Li_{0.03}O_4$)[23]. These theoretical and experimental results allow us to be assured that a doped hole in a $CuO_2$ layer can induce a weak ferromagnetic moment by reversing the average spins in a finite region of the antiferromagnetic layer around the dopant, as schematically shown in Fig. 1c. The above consideration is based not only on the theoretical investigations[12-22] where the theory is not about to break down but also on the experiments[23]. Thus, the polaron-based Skyrmion model shown in Fig. 1c can be a sound underlying basis to further move on to the higher hole density region. In what follows, we refer to the Cu atom surrounded by a single hole as Cu(*h*).

**Spin states of a doped $CuO_2$ plane**

We next investigate the possible spin sates of a $CuO_2$ plane in the underdoped and optimally doped regions. According to the Skyrmion-based picture mentioned above, the direction of their magnetic moment around the Cu(*h*) atom is determined by reversing that of the Cu atom in the original undoped $CuO_2$ plane. Since the doped hole is localized mainly on the O atoms around the Cu atom in the respective $CuO_4$ plaquettes[17,18,21,24], the neighboring Skyrmions will not highly modulate each others' magnetic moment as long as the distance between the two Cu(*h*) atoms is longer than a few lattice units.

When the hole concentration $p$ per Cu is quite low ($p<\sim0.05$), the configuration of the Cu($h$) atoms in the CuO$_2$ plane will not be uniquely determined. In such a low doping level, all Cu($h$) atoms will be maximally and randomly separated in the CuO$_2$ plane because of the mutual repulsive interaction[21] (see, for example, Fig. 2a). When the hole concentration reaches a value of $p=1/9$ (~0.11), each Cu($h$) atom will be located in the center of the 3×3 Cu supercell to realize maxim separation between the neighboring Cu($h$) atoms (see Fig. 2b). This creates symmetrical magnetic domains consisting of Skyrmion-derived ferromagnetic ordering regions, where the adjacent magnetic domains point in opposite directions, forming an $8a_0 \times 8a_0$ square unit. Accordingly, a magnetic pattern with antiparallel magnetic order is created at $p=1/9$. The present hole-driven magnetic ordering is expected to be energetically stable since the net magnetization and the resulting magnetic-energy density are zero. This allows us to expect that the spatial configuration of the Cu($h$) atoms in the CuO$_2$ plane with $p > \sim0.1$ tends to be governed not only by their repulsive forces but also by the total magnetic energy of the resulting magnetic pattern.

A similar but not exactly the same antiparallel magnetic order can be obtained for $p=1/6$(~0.167). This hole concentration almost corresponds to the optimal doping level showing a maximal superconducting transition temperature ($T_c$) for a number of cuprates. As for $p=1/6$, where one Cu($h$) atom is present in every 3×2 Cu supercell, it is possible to put the Cu($h$) atoms with keeping the C$_{2v}$ symmetry, forming one-dimensional ferromagnetic chains with $8a_0$-wide periodic ordering, as shown in Fig.

2c, The ferromagnetic chains with opposing magnetizations are alternately aligned, vanishing net magnetization. One may also construct other magnetic patterns consisting of more fragmented magnetic domains with no net magnetization for $p=1/6$. We should note, however, that there is a cost in energy involved in forming transition region between magnetic domains[25]. Thus, we consider that unidirectional antiparallel magnetic order, shown in Fig. 2c, is the most energetically favored configuration in view of the magnetic-energy density.

It is reasonable to expect that the spin state of the doped $CuO_2$ plane with $p$ ranging from ~1/9(~0.11) to ~1/6(~0.17) is characterized also by antiparallel magnetic order to minimize the total magnetic-energy density; for example, the spin state for $p=1/7$ (0.14) is expected to have a 1:1.5 mixture of the magnetic patters expected for the ideal $p=1/9$ and $p=1/6$ states. As for the slightly overdoped region ($p>$~0.18), however, such an ideal antiparallel magnetic order is difficult to be realized because of the presence of the intervening Cu($h$) atoms between the antiparallelly aligned ferromagnetic chains (see Fig. 2e). When the hole concentration exceeds ~0.25(~1/4), where every 2×2 Cu supercell contains more than one Cu($h$) atom, the ferromagnetic region of the neighboring Skyrmions begins to overlap, which induces magnetic frustration between the two Skyrmions (see Fig. 1d). Consequently, ferromagnetic ordering of the Skyrmions will be strongly diminished or vanished in the $CuO_2$ plane with $p >$~0.25.

**Model of the Cooper pair attraction**

In the preceding section, we have shown that the simple Skyrmion-based model yields the antiparallel magnetic pattern in the hole-doped $CuO_2$ plane. We should note that antiparallel magnetic order with a large spatial extent has been observed in the pseudogap region of cuprate superconductors[26,27]. Furthermore, it has been demonstrated that static magnetic order with spatial periodicity of $8a_0$ oriented along the Cu-O bond directions is present in and around the vortex cores of the cuprate superconductors[28-30]. This correspondence strongly implies that the antiparallel-magnetic-ordering state obtained here captures underlying spin states in the hole-doped $CuO_2$ systems. If the implication is valid, there should be a mechanism that can account for the formation of the superconducting state starting from the proposed ordering state.

In the $CuO_2$ planes, the doped holes reside primarily on O sites. The spin direction of the doped hole is opposite to that of the Cu($h$) $d$-hole because of a strong antiferromagnetic interaction between an O $p$-hole and a Cu $d$-hole[21,22] (see also Fig. 1). It hence follows that the spins of doped holes are aligned in an antiparallel manner according to the configuration of Cu($h$) atoms in the magnetically ordered $CuO_2$ planes (see Fig. 3a), forming a two-dimensional $S=1/2$ system. It should be noted once again that the doped $CuO_2$ plane itself also has antiparallel magnetic order, or antiferromagnetic-like order, consisting of a spin texture of Skyrmions. It is interesting to note that there exists a wide range of antiferromagnetic systems that can be turned to a superconducting state on the border of long range antiferromagnetic order[31-33], which is known as magnetically mediated superconductivity[31,34] or spin-fluctuation-mediated superconductivity[32]. Thus, we suggest that spin singlet paring between the two adjacent holes will be mediated by antiparallel magnetic order in the hole-doped $CuO_2$ plane. If the pairing mechanism indeed occurs, the spatial periodicity of the charge density will be reduced to one-half of the initial magnetic domains with $8a_0$-periodic modulation (see Fig. 2d and Fig. 3b). Accordingly, the coherence length of the Cooper pair parallel to the $CuO_2$ layer $\xi_{ab}$ is on the order of this modulation ($4a_0$~1.5 nm), which is in reasonable agreement with a typical value of $\xi_{ab}$ (~1.4 ± 0.2 nm) reported for cuprate superconductors[35].

It should also be worth mentioning that the assumed pairing scheme results in a sub-nanoscale superconductor/ferromagnet (S/F) hybrid multilayer, in which each superconducting domain is sandwiched by two ferromagnetic domains with opposing magnetizations although the boundaries may not be well defined (see Fig. 3b). In such an S/F hybrid system, the proximity effect[36,37] is expected to occur. That is, the Cooper pairs can penetrate into the F layers and induce superconductivity there; the order-parameter normally shows an oscillatory decay in the F layer[36]. In the case of an S/F multilayer, the oscillations may vary depending on the thickness of the S and F layers[36]. The 0 phase corresponds to the same superconducting order-parameter sign in all S layers (see Fig. 3c), while in the π phase the sign of the superconducting order-parameter in adjacent S layers is opposite (see Fig. 3d). Thus, the case of the 0 phase is likely to be consistent with the *d*-wave pairing state in cuprate superconductors.

Previously, a possible correlation between antiferromagnetic spin fluctuation and high-temperature superconductivity has actually been proposed[38,39]. It has also been demonstrated that the observed *d*-wave pairing is fully compatible with the antiferromagnetic spin fluctuation[34,38,39]. However, one of the obstacles against the antiferromagnetic spin fluctuation scheme is the presence of a pseudogap state[38], which lies in between the antiferromagnetic and superconducting states in a generic phase diagram of hole-doped cuprates. Our scheme, however, is not subject to this problem because the antiferromagnetic state mediating the superconducting state is not the Mott insulating state in the undoped $CuO_2$ plane but the one created by hole doping. According to our scheme, the superconducting transition temperature $T_c$ can be viewed as a limiting temperature at which the long-range antiferromagnetic order in the doped $CuO_2$ plane emerges to yield a phase-coherent supercurrent[40]. When the temperature of the system rises above $T_c$, the long-range antiferromagnetic order will be partially broken because of the thermal fluctuation of the magnetic domain, leading to the loss of long-range phase coherence and the disappearance of the superconducting state. However, this does not necessarily mean that Cooper pairs do not exist at all at temperatures above $T_c$. It is probable that the antiferromagnetic order still survives above $T_c$ at least on the medium-range length scale (up to several tens of lattice units), giving rise to phase-incoherent pairing. Although this partially ordered state will not contribute to the phase-coherent supercurrent, it is reasonable to expect that this state can affect various electronic and magnetic properties, such as the density of states near the Fermi level, the real-space charge distribution, vortex states under external magnetic field, and the electrical conductivity, etc. We hence propose that the medium-range antiferromagnetic order with the related pair condensate underlies a number of complex behaviors observed in the pseudogap state[41], for example, stripe and checkerboard charge ordering[30,42], unusual magnetic order[26,27], and a large enhancement of the Nernst signal[43]. It is clear from Fig. 2 b,c that the assumed hole-driven magnetic/charge states inherently have stripe and checkerboard patterns.

With a further increase in temperature, the hole-driven antiferromagnetic order will eventually and completely be lost at a certain temperature $T^*$, possibly entering another thermodynamic phase. Although any antiferromagnetic order will not exist above $T^*$, transport properties of the doped holes are severely affected by the strong electron correlation effect in the system as long as the hole concentration is not so high as to suppress the effect. This probably accounts for the occurrence of the 'strange-metal' region or the non-Fermi liquid state.

**Interpretation of the 1/8 anomaly**

Lanthanum '214' cuprates, for example, $La_{2-x}Ba_xCuO_4$, show a deep depression of $T_c$ and the highest stripe ordering temperature at $p \sim 1/8$[44-49]. In previous sections, we have shown that the $CuO_2$ planes with $p=1/9$ and $1/6$ have characteristic antiferromagnetic modulations, leading to the Cooper pair formation. It is hence interesting to investigate what will happen to the $CuO_2$ plane with $p=1/8$ in terms of the hole-driven magnetic order.

As for $p=1/8$, we can put the $Cu(h)$ atoms in the $CuO_2$ plane with keeping the $C_{2v}$

symmetry, as shown in Fig. 2f, where one Cu(*h*) atom is present in every 4×2 Cu supercell. The resulting $CuO_2$ plane has one-dimensional ferromagnetic chains, similar to the case of *p*=1/6. In the $CuO_2$ plane with *p*=1/8, however, the neighboring ferromagnetic chains point to the same direction to yield stripe ferromagnetic order. On a macroscopic lengths scale, stripe regions pointing in opposite magnetic directions (see Fig. 2f) may be coexisting to reduce the total magnetization-energy density. We should also note that the spin and charge stripe pattern shown in Fig. 2f is different from the previously proposed one[44], where the orientation of magnetic moment on Cu atoms are locally antiparallel and the spin direction rotates by 180° relative to a simple antiferromagnetic structure on crossing a domain wall on which the doped holes are located. According to our pairing scheme, the spin singlet paring never occurs in the ferromagnetic regions shown in Fig. 2f because of the lack of long range antiferromagnetic order. This accounts for a deep depression of $T_c$ at *p*=1/8 observed for the lanthanum '214' family. Although any pairing cannot be expected, unidirectional ferromagnetic order shown in Fig. 2f may more or less survive at higher temperatures. Thus, the robust stripe ordering at *p*=1/8 can be naturally interpreted in terms of our magnetic-domain-based model, also providing an indirect support to our pairing mechanism.

**Conclusions**

On the basis of the above considerations, we can summarize and rationalize the generic feature of a phase diagram of hole-doped cuprates, as shown in Fig. 4.

In a very lightly doped region ($p<\sim0.05$), the Cu($h$) atoms with Skyrmion-type ferromagnetic order are scattered and randomly distributed in the $CuO_2$ plane, as shown in Fig. 2a and Fig. 4a. When the motion of these randomly distributed doped holes is freezed at a certain temperatures, the resulting $CuO_2$ plane will show a spin-glass freezing behavior. Thus, the spin glass phase seen in a very lightly doped region can be viewed as a disordered magnetic state due to hole-doping driven ferromagnetism (HDF).

Upon increasing doping level $p$ from $\sim0.1(\sim1/9)$ to $\sim0.17(\sim1/6)$, the antiparallel magnetic pattern, or hole-doping driven antiferromagnetism (HDAF), tends to prevail. When long-range phase coherence is established at $T_c$, the system enters a magnetically mediated superconducting state. The resulting superconducting phase can hence be viewed as long range ordering state of HDAF. At temperatures above $T_c$, ordering of HDAF will be preserved only on the length scale of several tens of lattice units, giving rise to the phase-incoherent paring state responsible for the pseudogap phase. With further increase in temperature above $T^*$, any magnetic order derived from HDAF will be lost, transforming to a non-Fermi liquid phase.

When the hole concentration exceeds $\sim0.25(\sim1/4)$, i.e., ferromagnetic ordering of the Skyrmions will be virtually vanished because of the magnetic frustration between the neighboring Skyrmions (see Fig. 4e), leading a normal Fermi liquid state.

Thus, the apparently unavoidable complexity recognized in cuprate superconductors can be reasonably understood in terms of the hierarchy of ordering pattern of HDAF. Thus, HDAF can be regarded as a 'parent'[50] state out of which the many types of orders, including superconducting order and charge- and spin-density wave orders, emerge. We believe the present model provides a simple but essential component to construct a complete theoretical framework for full understanding of high-temperature cuprate superconductors.

**Figure 1. Schematic representation of a hole in the CuO$_2$ layer.** Large and small circles represent O and Cu atoms, respectively. **a**, Zhang-Rice singlet state[8]. **b**, Ferromagnetic moment induced via Skyrmion formation, **c**, Simplified description of the hole surrounding the central Cu atom in a CuO$_4$ plaquette and its neighboring Cu atoms. The shaded region represents the expected Skyrmion-type spin texture. **c**, Frustration of magnetic moment between the two adjacent Skyrmion, diminishing the Skyrmion-derived ferromagnetic moment.

**Figure 2. Variations of the magnetic domains in the CuO$_2$ plane upon hole doping.** The hole concentration per Cu $p<\sim0.05$ (**a**), $p=1/9$ (**b**), $p=1/6$ (**c**), $p>1/6$ (**e**), $p=1/8$ (**f**). The oxygen atoms, which surround the Cu sites, have been omitted. The two-headed arrows represent the spin on the Cu atom surrounded by a doped hole, Cu($h$). Domains with the same magnetic directions are shaded by the same colors: red and blue shaded regions represent the upper and down spin regions, respectively. $a_0$ corresponds to the Cu-O-Cu bond distance. **d**, Schematic representation of magnetically mediated pairing state in the CuO$_2$ plane with $p=1/6$.

**Figure 3. Magnetically mediated paring mechanism in the hole doped CuO$_2$ plane.** Arrangement of spins aligned along the direction perpendicular to the rotational

axis of the hole-doped $CuO_2$ plane shown in Fig. 2b,c. Spin on Cu(*h*) is represented by a two-headed arrow. A linear antiferromagnetic chain consisting of doped hole spin is represented by a series of arrows above the Cu(*h*) atoms. **a,** Before pair formation. **b,** After pair formation. Schematic representation of the superconductor(S)/ferromagnet(F) multilayer is also shown. **c,** Schematic behavior of the superconducting order parameter in the superconductor(S)/ferromagnet(F) multilayer. The curve $\Psi(x)$ represents schematically the behavior of the Cooper pair wavefunction in the 0 phase. Due to symmetry the derivative of $\Psi(x)$ is zero at the center of the S and F layers. **d,** The Cooper pair wave function in the $\pi$ phase, where $\Psi(x)$ vanishes at the center of the F layer.

**Figure 4. Generalized phase diagram of hole doped cuprate materials. a-d**, Doped-hole driven magnetic pattern; $p<\sim 0.05$ (**a**), $p=\sim 0.111(1/9)$ (**b**), $p=0.125(1/8)$ (**c**), $p=\sim 0.167\ (1/6)$ (**d**), $p>\sim 0.25$. DHAF and DHF denote doped-hole driven antiferromagnetic and doped-hole driven ferromagnetic regions, respectively.

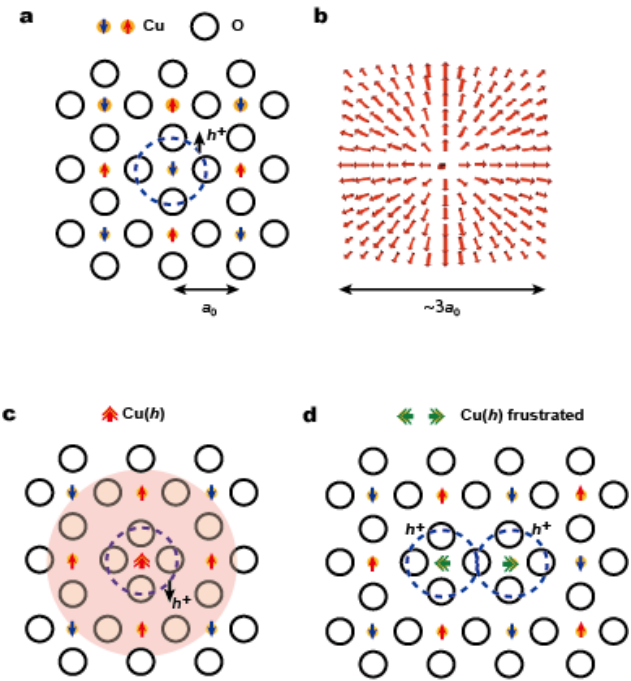

Fig. 1

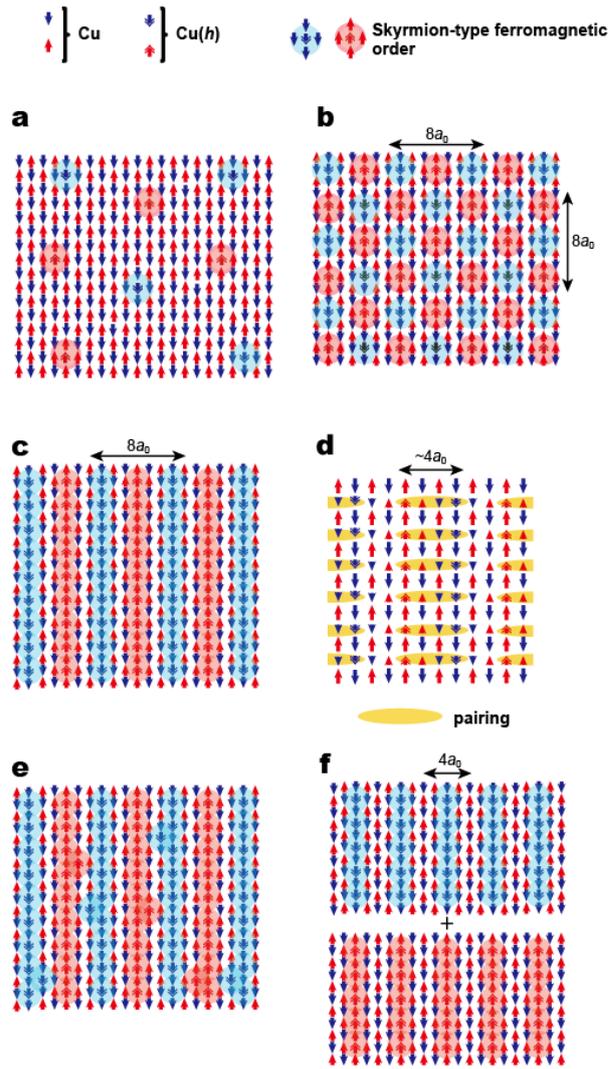

Fig. 2

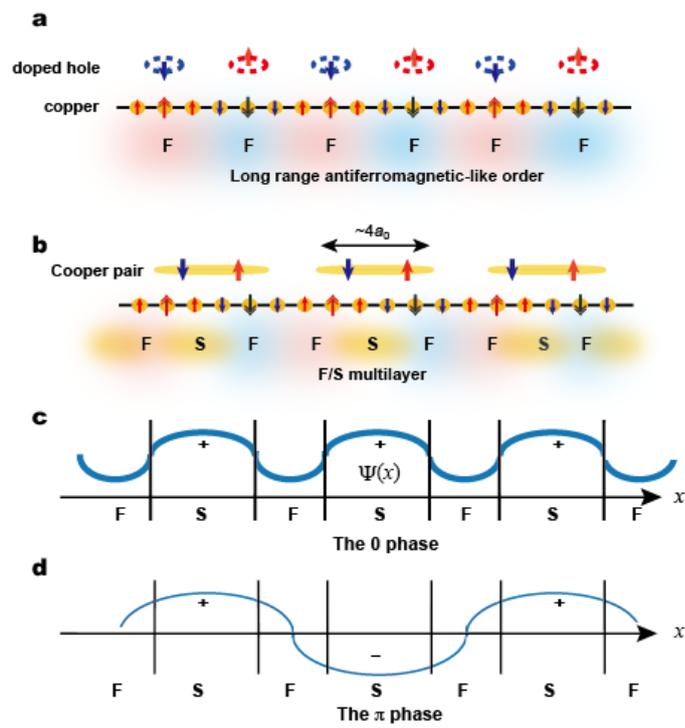

Fig. 3

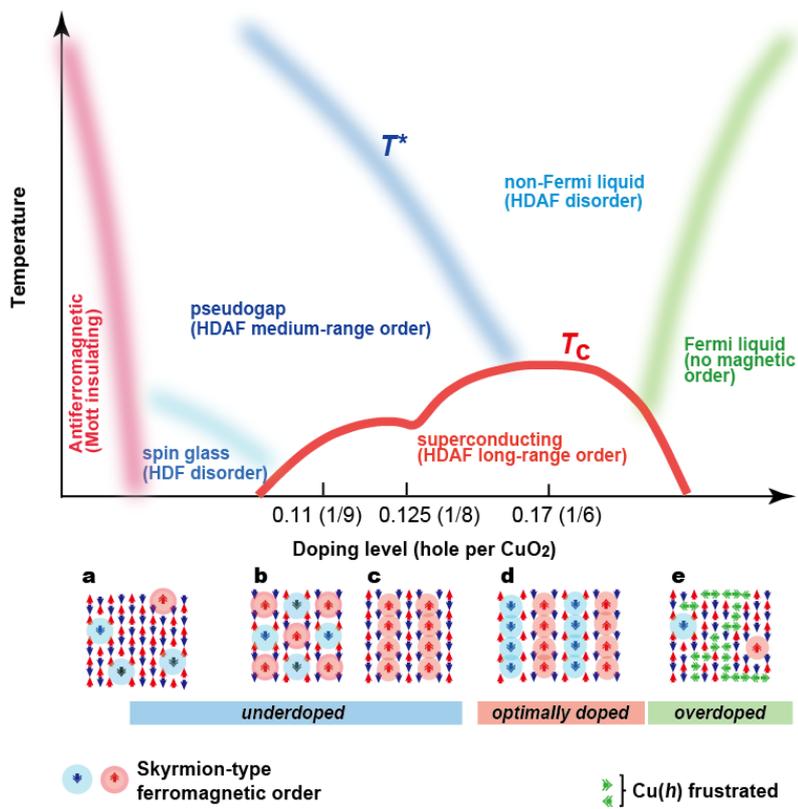

Fig. 4